# Holding the Cosmos in Your Hand: Developing 3D Modeling and Printing Pipelines for Communications and Research

**Kimberly K. Arcand[1], Sara R. Price[1], Megan Watzke[1]**

[1]Chandra X-ray Observatory, Center for Astrophysics | Harvard & Smithsonian, Cambridge, MA, United States

**\*Correspondence:**
Sara R. Price
sara.price@cfa.harvard.edu

**Keywords: 3D printing, 3D visualization, virtual reality, astrophysics, geophysics, science communication, inclusivity**

**Abstract**

Three-dimensional (3D) visualization has opened up a universe of possible scientific data representations.  3D printing has the potential to make seemingly abstract and esoteric data sets accessible, particularly through the lens of translating data into forms that can be explored in the tactile modality for people who are blind or visually impaired. This article will briefly review 3D modeling in astrophysics, astronomy, and planetary science, before discussing 3D printed astrophysical and planetary geophysical data sets and their current and potential applications with non-expert audiences. The article will also explore the prospective pipeline and benefits of other 3D data outputs in accessible scientific research and communications, including extended reality and data sonification.

## 1 Introduction

This paper will summarize recent achievements made in astronomy, astrophysics, and space science using three-dimensional (3D) technology to help reach beyond expert audiences, and discuss best practices that can be shared with other sciences such as geophysics. This type of 3D work also allows for new media -- both virtual and in tangible physical form -- to help non-experts interact with and learn about these discoveries. We contend that the lessons learned from these space-based 3D projects could have broader impacts for many fields throughout science, technology, engineering, and mathematics (STEM) fields and provide educational benefits to a great number of people.

Parallel advances in acquiring 3D data from Earth and space and in 3D printing mean that today technologists can translate digital data into something that we can touch and feel, expanding the ways we can understand and communicate science, including with diverse communities such as the blind or visually impaired (BVI) (Trotta et al., 2020). Currently, the amount of astronomical and planetary geological objects able to be 3D printed is low compared to the vast range of objects being studied (Arcand et al., 2017). One practical challenge is making 3D modeling easier and more accessible as a research tool. This issue can be improved with the advent of enhanced data from the next generation of more technically advanced telescopes and sensors (Gerloni et al., 2018). Creating awareness among scientific researchers of 3D data representation is also essential to increasing this technique's frequency and convenience of use.

This article describes recent innovative 3D processes and explores the developing pipelines and contributions of 3D visualization outputs in multimodal scientific study and communications. 3D printing, for example, is a potential place of interaction between terrestrial and space science through representations of discipline-specific and interdisciplinary data alike, aimed at both non-experts and experts that can use specific accessible outputs and products. Indeed, best practices for 3D printing of scientific data can and should be shared across traditionally separate disciplines.

Astronomical and other scientific information can be shared through various human senses including touch, smell, and even taste (Trotta, 2018). The use of, and meaning-making from, scientific visualizations and related expressions of data "depend strongly on who is viewing" (Steffen, 2016, para.4) or interpreting them. Consequently, accessibility and diversity are a critical area of emphasis for this topic, and we will summarize current evidence of and future needs for the impact of 3D visualization on enhancing equity via modeling and printing. Our analysis briefly encompasses extended reality (XR) - a suite of multisensory applications which typically involves augmented, virtual, and mixed reality formats (Najjar, 2020) for diverse communities of users.

The COVID-19 pandemic has shown the potential for 3D modeling and printing – from making personal protective equipment to respirators[1] and ventilators (e.g., Molitch-Hou, 2020) to understanding the structure and impact of the coronavirus through 3D printing (Grove, 2020) and virtual reality (VR) (e.g., Pearson, 2020). Although geophysics and astrophysics data typically cover very different phenomena, they can benefit from multidimensional data delivery to diverse audiences during times of disruption and beyond.

## 2      3D Visualizations & Related Outputs

Learning about the mental manipulation of 3D images is an important step in the process of science education in both astronomy (Eriksson et al., 2014) and geology (Carneiro et al., 2018). 3D modeling skills can form the basis of a framework for teaching astronomy and related topics in a manner that enables non-experts to use and understand the same visual and conceptual assets as professional astronomers (Eriksson, 2019). Multidimensional models are also useful as teaching tools in geophysics to share the textures and geological arrangement of surfaces including sedimentary basins with students and the public (Carneiro et al., 2018). Simultaneously, the ability to research celestial or geophysical objects from multiple angles and viewpoints can help improve understanding of how scientific objects are structured or how the underlying physics works (Ferrand & Warren, 2018), making them a valuable tool. Once a 3D visualization has been created, there are multiple output potentials from interactive 3D digital models or physical 3D prints to immersive XR experiences (Arcand et al., 2018). 3D printing and other outputs involving touch can help encourage more "control" (Isenberg, 2013) and interactivity on behalf of users (Madura, 2017) than computerized 3D visualization usually allows.

3D printing is a process where a type of material such as plastic, metal, or an organic substance is continually added layer by layer to formulate the object (Grice et al., 2015). Recently, "on-demand processes" of 3D printing -- typically characterized as fused deposition modeling -- have become increasingly accessible and affordable for consumers (Takagishi & Umezu, 2017), including those in libraries and related educational and community settings. This opens the opportunity to produce scientific or educational tools in these learning environments that may have been unachievable before (Hasiuk et al., 2017). On consumer scales, 3D printing is still rather new, but the production possibilities are far-reaching. 3D printing applications in science range from plans for an on-demand and sustainable lunar base 3D printed from Moon dust (De Kestelier et al., 2015) to medical 3D printing of skin cells to help treat burn victims (Brown, 2017; Everett-Green, 2013).

---

[1]  https://medeng.jpl.nasa.gov/covid-19/respirators/, accessed 26 June 2020

The prominence of VR as an entertainment product in recent years has made such technologies more widely available for scientific communications and research in astronomy as well (Baracaglia & Vogt, 2020). Due to the extensive time and technical effort often needed to develop VR applications for astronomical or geological objects, modifiable generic VR templates could be helpful for the scientific community, and could help spark greater adoption for observational exploration (Baracaglia & Vogt, 2020). VR is another 3D tool that can potentially be used by scientists for visualizing the immense amount of "big data" making up modern astronomy, in particular for data mining, detection, removal, and quick-look rendering (Comparato et al., 2007; Donalek et al., 2014). Interactive VR is also able to take advantage of human depth perception as an ergonomic factor that enables quick identification and characterization of complex structures (Steffen, 2016). Augmented reality (AR), which overlays digital elements onto a user's perception of the surrounding world, and VR can be important support tools in the processing of massive and complex scientific data sets. However, for any XR product, individual differences in the way people experience multimodal information should be taken into account (Olshannikova et al., 2015).

## 2.1     Importance of Audience in Scientific Visualization: Inclusivity and Diversity

Astronomy and astrophysics (Rector et al., 2015) and geology and geophysics (Pound, 2019) are often recognized as highly visual fields both historically and today. This can create challenges in sharing data from these disciplines with BVI audiences. Evaluative studies have shown benefits in using 3D astrophysical models (Arcand et al., 2019; Argudo-Fernández et al., 2020; Bonne et al., 2018; Christian et al., 2015; Grice et al., 2015) for generating or positively impacting learning gains, inclusive practices, STEM identity, and mental visualization. Applications of 3D models in geology and geophysics have been demonstrated for use in museums and similar informal learning environments (Neitzke Adamo et al., 2019) and by other educators (Hasiuk et al., 2017).

In the past several years, research has shown that 3D printed scientific data from astrophysics and geophysics with tactile features can help communicate with BVI participants across a spectrum of abilities (Bonne et al., 2018) as well as with sighted people (European Southern Observatory, 2019), and also to promote inclusivity more broadly (Arcand et al., 2019; Christian et al., 2015). XR technologies particularly have been broadly shown to provide low-risk, high-impact virtual spaces that can accommodate physical barriers to interaction by providing learner-specific experiences (Chandrashekar, 2018), when crafted particularly through universal design techniques (McMahon & Walker, 2019; Menke et al., 2020).

Human and interpersonal issues must be considered in order to enable meaningful use of 3D prints and dynamic virtual spaces for science communication with BVI and other audiences. Audience awareness and the resulting specific development of visualizations with audience needs in mind is critical for the development of inclusive practices in 3D visualizations, facilitated by 3D printing and other emerging technologies (Hurt et al., 2019; see e.g., Diaz-Merced et al., 2011; Diaz-Merced, 2013; Diaz-Merced, 2014; Christian et al., 2015; Grice et al., 2015; Madura, 2017; Madura et al., 2015; Steffen et al., 2011; Steffen et al., 2014).

Across the United States (ADA National Network, 2019; Smithsonian Institution, 2014) and around the globe (United Nations — Disability, Department of Economic and Social Affairs,

2006), there are firm legal requirements to provide and maintain access to information, communication, and participatory opportunities for people with disabilities. Differently-abled populations need to be able to discover and share information in a way directly equivalent to how others complete such tasks (United Nations — Disability, Department of Economic and Social Affairs, 2006). People with visual impairments, for example, exist with a specific spectrum of needs that can be affected by variables such as the timing and cause of their blindness' onset (National Federation of the Blind & Finkelstein, 1994) or the amount of Braille literacy instruction (National Federation of the Blind Jernigan Institute, 2009).

Therefore, any effort should take into account how to effectively disseminate 3D prints and related Braille or tactile materials. Incurred costs can be problematic for organizations and individuals (Arcand et al., 2019; Beck-Winchatz & Riccobono, 2008; Weferling, 2007) as well as intended audiences, including BVI communities. These issues are particularly relevant during times of disruption and upheaval, as illustrated during the COVID-19 pandemic (McEvoy, 2020).

# 3      Key Examples of 3D Visualizing in Planetary Geophysics and Astrophysics

Planetary geology and geophysics are particularly well suited areas of study for the development of 3D visualization outputs including 3D models, 3D prints, and XR. 3D visualizations of the Earth can clarify geological phenomena to the public (Kyriakopoulos, 2019), while also supplementing instruction (Koelemeijer et al., 2019) and making it more accessible (Dolphin et al., 2019). In exoplanet research, studies on the habitability of exoplanet bodies like the TRAPPIST-1 system[2] (Jet Propulsion Laboratory, 2020; NASA, 2019), and synthetic exoplanet systems (Alesina et al., 2019) have led to head-set based VR experiences for expert and non-expert audiences. 3D models have been useful in the study of asteroids (Kim, 2018) and aspects of other planetary bodies such as the atmospheres of Venus (Korycansky et al., 2002) and Titan (Charnay et al., 2014) as well as landscape evolution (Cornet et al., 2017).  A wealth of data on our nearest neighbors, our Moon and Mars, has resulted in extensive 3D mapping of the various local topographies and geological structures (Edwards et al., 2005; Ellison, 2014; Gwinner et al., 2015; Löwe & Klump, 2013; Mars Exploration Rovers, 2014), and morphometric globes (Florinsky & Filippov, 2017) of these objects with outputs of 3D visualizations, primarily for scientific analysis.

In astronomy and astrophysics, data and simulation driven 3D printed objects include multiple object types across a range of scales. In high-energy astrophysics, supernova remnants have been particularly conducive to 3D modeling, encompassing such examples as Cassiopeia A (Arcand et al., 2017; Arcand et al., 2018; Arcand et al., 2019; DeLaney et al., 2010; Orlando et al., 2016), Tycho's supernova remnant (Chandra X-ray Observatory, 2019; Ferrand et al., 2019), Supernova 1987a (Arcand et al., 2017; Arcand et al., 2019; Orlando et al., 2018), and the Crab Nebula (Arcand et al., 2020; Summers et al., 2020). Beyond supernovae, other areas of 3D research and output range from star formation regions such as M16 (McLeod et al., 2015), a massive star system with colliding winds called Eta Carinae (Arcand et al., 2017; Madura et al., 2015; Madura, 2017; Steffen et al., 2014), double star system V745 Sco (Arcand et al., 2019; Chandra X-ray Observatory, 2017), protostellar jets like DG Tau (Chandra X-ray Observatory, 2020a; Orlando et al., 2019; Ustamujic et al., 2016; Ustamujic et al., 2018), and beyond to much larger structures including the South Pole Wall (Pomarède et al., 2020), the

---

[2] http://www.spitzer.caltech.edu/vr; accessed 17 June 2020

Cosmic Web (Diemer & Facio, 2017) and the Cosmic Microwave Background (Arcand et al., 2017; Clements et al., 2016).

Multidimensional renderings of astronomical and geological data in accessible formats can simplify the discovery of previously hidden or overlooked structures in objects and, through the presence of interactive features, can enable close-up views of data via a personalized perspective (Madura, 2017). This variety of technologies that have proven useful for accessible public engagement has helped to clarify the context of current observational science data (Arcand et al., 2017; Madura et al., 2015). The catalog of data-driven astrophysical or geological models continues to grow (Hurt et al., 2019), and can be used in communicating contemporary science with non-expert audiences (Löwe & Klump, 2013). Additionally, artistic imaginings with scientific underpinnings (Pauwels, 2020), in contrast to data-driven or simulation based 3D models, can tangibly provide perceptual value in communicating information that can be abstract, esoteric, or perhaps invisible to the human or robotic eye (Keefe et al., 2005) particularly in this area of multidimensional visualization.

One important factor for the accessibility of 3D models, particularly with non-experts, is the availability of open access or Creative Commons data. The National Aeronautics and Space Administration (NASA)[3] and the Smithsonian Institution (Pearlman, 2020), for example, each maintain open access or public domain databases of 3D objects, ranging from supernova remnants to geological maps of the Apollo lunar landing sites. Open access and Creative Commons materials can reduce complexities or legalities regarding the adaptation of content to customized learning experiences for multiple audiences (Zhang et al., 2020). Flexible, customizable materials, derived from openly accessible materials, can potentially improve experiences for all learners, particularly for those with special access requirements (Zhang et al., 2020).

## 4      Discussion

### 4.1    3D Visualization Pipelines

Once a 3D visualization has been created, there are myriad options for how to output that data into a usable product, specific to an intended audience. This underlines the importance of understanding and establishing pipelines in creating 3D data sets, whether ultimately working with file types from .vtk to .obj to .stl. to .unity3d or beyond, for scientific research or for scientific engagement.  Narrative information should be provided to interpret context and scale.

### 4.2    XR Technology Examples Related to the Data Pipeline

Preliminary astrophysical VR applications of simulated worlds (such as Farr et al., 2009) through recent astronomical VR experiences as individual applications (e.g., Arcand et al., 2018; Chandra X-ray Observatory, 2020b; Ferrand & Warren, 2018; Russell et al., 2017), include artistically illustrated worlds, simulated data mapped to astronomical observations, and three-dimensional models derived from scientific observations. Users can comb Martian surfaces based on observational data from NASA's Jet Propulsion Laboratory (2017), travel to the TRAPPIST-1 exoplanets via scientifically-influenced 3D artists' impressions that were converted into VR applications (NASA, 2019), and explore the solar surface based on observational data from the Hinode spacecraft (Hinode Science Center at NAOJ, 2018). Explorers equipped with emerging technologies can also virtually walk inside the remains of an exploded star (Arcand et al., 2018) or experience a spiral galaxy like NGC 3198 via radio data cubes (Ferrand et al., 2016). Computational models and simulations constrained by scientific observations also provide numerous options for VR application development, including simulation and time-domain based applications of Sagittarius A, the supermassive black hole at our Galactic Center (Davelaar et al., 2018; Russell, 2017).

---

[3]   https://nasa3d.arc.nasa.gov/models, accessed 16 June 2020

Expert analysis of scientific data in XR spaces can provide active and immersive simulations of astrophysical topics like photometry of the billions of stars in our Milky Way (Ramírez et al, 2019). And at even greater scales, cosmological topics such as dark matter can be rendered in VR with particular attention towards accessibility, for example haptic (vibrational) cues that work together with wheelchair use (Aviles, 2018). Current research into, and evaluation of, such immersive projects that take advantage of human perception in 3D spaces are investigating potential outcomes in enabling better detections and manipulations of large or complex data sets on behalf of the scientist (Ferrand et al., 2016). Additionally, in geology and geophysics there are projects across topographical maps (Woods et al., 2015), geological surveys (Westhead et al., 2013) and other ways to provide 3D data in a digestible way to "enable non-expert users to begin to interact with a complex science in an expert way which was not possible for previous generations" (Westhead et al., 2013, p.189; see e.g, Gerloni et al., 2018; Mathiesen, 2012; Trexler et al., 2018).

Data sonification is another example of a potential extension or output of 3D data sets. Sonification can help improve upon analysis of big data through vision alone by taking advantage of the unique capacities of sound to provide information from different dimensions simultaneously, with quick interaction or playback (Cooke et al., 2017). This strategy can work with XR in order to speedily evaluate features of 3D data (Ribeiro et al., 2012). Diaz-Merced (2013) demonstrated that listening function can be improved through targeted interventions as applied to such data sonification outputs. Beyond data sonification there are additional 3D visualization outputs and enhancements that can be considered to reach specific audiences, from holograms (Royal Astronomical Society, 2019) and haptic information (Isenberg, 2013; Trotta et al., 2020) to tactile tablets (Touch Graphics, Inc., 2015) and multisensory experiences (Najjar, 2020).

## 5      Conclusion

At this time, the COVID-19 pandemic continues to result in monumental challenges that reach nearly every aspect of life. The pandemic does not spare the fields of science research, visualization or engagement, but instead presents particular obstacles -- and opportunities – to these areas. Some of the recent difficulties related to the use of 3D printed materials include reliance on tactile surfaces while combating a virus that can spread through surface contact (Centers for Disease Control and Prevention, 2020) and a reduction in tangible resources presently available to populations who need enhanced accessibility, such as BVI audiences (McEvoy, 2020). However, due to social distancing there is now a greater demand than ever before in modern times for remote learning and work. The 3D astrophysical and geophysical models, prints, and virtual spaces could serve scientists, educators, learners and others who are unable to spend their usual time in physical settings. We propose partnering with groups who both advocate for and engage with affected audiences so the science and engagement communities can best serve the needs of their audiences.

**Author Contributions**

KA, as the principal investigator, provided the main points, literature references and scientific topics, as well as technical writings for the article.

SP, as second author and research assistant, formatted the outline and citations, organized the literature review, and provided summative text on specific subsections.

MW, as third author, provided detailed editing and drafting, gave input on overall organization, and wrote summary texts as needed.

## Funding

This paper was written with funding from NASA under contract NAS8-03060 with the authors working for the Chandra X-ray Observatory. NASA's Marshall Space Flight Center manages the Chandra program. The Smithsonian Astrophysical Observatory's Chandra X-ray Center controls science and flight operations from Cambridge and Burlington, Massachusetts.

## Conflict of Interest Statement

The authors **declare** that the research was conducted in the absence of any commercial or financial relationships that could be construed as a potential **conflict of interest**.

## Acknowledgements

The Authors gratefully acknowledge Lisa Frattare for her stellar copy editing skills. The Authors also acknowledge the Chandra communications and public engagement group members, NASA's Universe of Learning, and the many scientists, technologists and researchers that have contributed to the ever growing library of 3D printed objects of our Universe.